\documentclass{article}

\usepackage{microtype}
\usepackage{graphicx}
\usepackage{subfigure}
\usepackage{booktabs}
\usepackage{hyperref}

\usepackage[accepted]{icml2020}

\newcommand{\resnet}{{ResNet-18}~}

\defcitealias{EHT1}{EHTC~I}
\defcitealias{EHT2}{EHTC~II}
\defcitealias{EHT3}{EHTC~III}
\defcitealias{EHT4}{EHTC~IV}
\defcitealias{EHT5}{EHTC~V}
\defcitealias{EHT6}{EHTC~VI}

\icmltitlerunning{Feature Extraction on Synthetic Black Hole Images}

\begin{document}

\twocolumn[
\icmltitle{Feature Extraction on Synthetic Black Hole Images}

\icmlsetsymbol{equal}{*}

\begin{icmlauthorlist}
\icmlauthor{Joshua Yao-Yu Lin}{phys}
\icmlauthor{George N.~Wong}{phys}
\icmlauthor{Ben S.~Prather}{phys}
\icmlauthor{Charles F.~Gammie}{phys,astro,icasu}
\end{icmlauthorlist}

\icmlaffiliation{phys}{Department of Physics, University of Illinois at Urbana-Champaign, U.S.A}
\icmlaffiliation{astro}{Department of Astronomy, University of Illinois at Urbana-Champaign, U.S.A}
\icmlaffiliation{icasu}{Illinois Center for Advanced Studies of the Universe}

\icmlcorrespondingauthor{Joshua Yao-Yu Lin}{yaoyuyl2@illinois.edu}
\icmlcorrespondingauthor{Charles F.~Gammie}{gammie@illinois.edu}

\icmlkeywords{Machine Learning, ICML}

\vskip 0.3in
]

\printAffiliationsAndNotice{}

\begin{abstract}

The Event Horizon Telescope (EHT) recently released the first horizon-scale images of the black hole in M87.  Combined with other astronomical data, these images constrain the mass and spin of the hole as well as the accretion rate and magnetic flux trapped on the hole.  An important question for EHT is how well key parameters such as spin and trapped magnetic flux can be extracted from present and future EHT data alone.  Here we explore parameter extraction using a neural network trained on high resolution synthetic images drawn from state-of-the-art simulations.  We find that the neural network is able to recover spin and flux with high accuracy.  We are particularly interested in interpreting the neural network output and understanding which features are used to identify, e.g., black hole spin.  Using feature maps, we find that the network keys on low surface brightness features in particular.

\end{abstract}

\section{Introduction}
\label{submission}

The Event Horizon Telescope (EHT) is a globe-spanning network of millimeter wavelength observatories~\citepalias{EHT1,EHT2}.  Data from the observatories can be combined to measure the amplitude of Fourier components of a source {\em intensity} on the sky~\citepalias{EHT3}.  The sparse set of Fourier components, together with a regularization procedure, can then be used to reconstruct an image of the source~\citepalias{EHT4}.  The resulting images of the black hole at the center of M87 (hereafter M87*) exhibit a ringlike structure---attributed to emission from hot plasma surrounding the black hole---with an asymmetry that contains information about the motion of plasma around the hole~\citepalias{EHT5, EHT6}.  Combined with data from other sources, the EHT images constrain the black hole spin and mass, as well as the surrounding magnetic field structure and strength~\citepalias{EHT5, EHT6}.

The EHT is now anticipating observations with an expanded network of observatories.  What can be inferred about the physical state of M87* from the improved data?  In particular, will it be possible to accurately infer the black hole spin (one of two parameters describing the black hole) from the EHT data alone?  And will it be possible to accurately infer the magnitude of the magnetic flux trapped within the black hole, which can have a profound effect on the interaction of the hole with the surrounding plasma and possibly launch relativistic jets?

\begin{figure}[t]
\vskip 0.2in
\begin{center}
\includegraphics[width=\columnwidth]{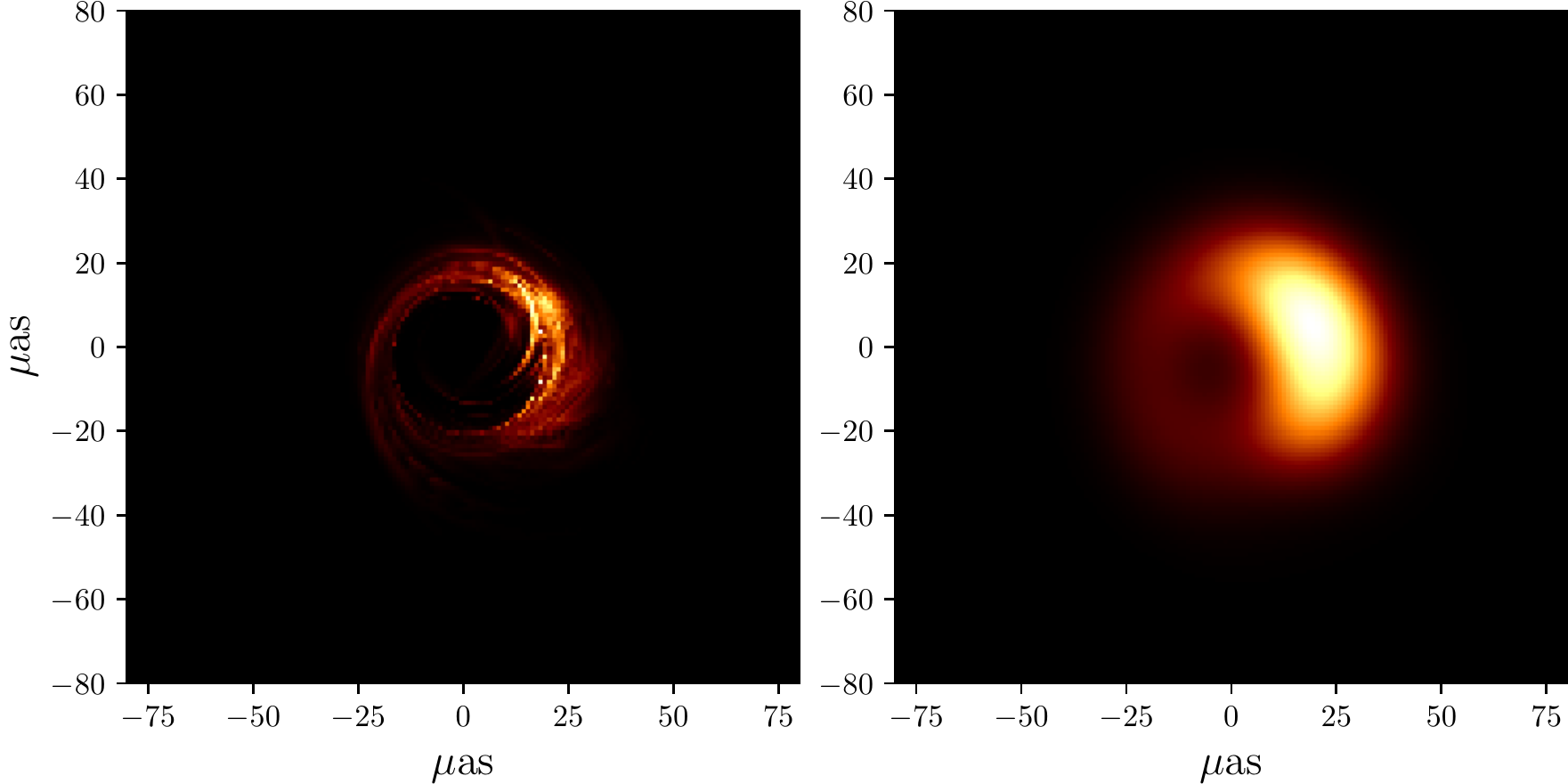}
\caption{Synthetic image examples. Left panel: full resolution simulated black hole image based on numerical model of black hole accretion flow with M87-like parameters. Right panel: the same simulated image convolved with $20 \mu$as FWHM Gaussian beam meant to represent the resolving power of the EHT.
}
\label{fig:Synthetic-black-hole-image}
\end{center}
\vskip -0.2in
\end{figure}

Interpretation of EHT data has been aided by magnetized, relativistic fluid models of the source, coupled to radiative transfer models.  These models enable the generation of large synthetic data sets that include the effects of black hole spin on the near-horizon plasma and electromagnetic fields, as well as emission and absorption of radiation by the synchrotron process. An example synthetic image is shown in Figure~\ref{fig:Synthetic-black-hole-image} at both high resolution (left) and EHT resolution (right).

Earlier work by~\citet{van2020deep} found evidence that neural networks could be used to infer the values for a limited set of parameters in a library of synthetic black hole images.  For the work presented here we have generated a special set of $10^6$ synthetic images.  The images are generated from a set of $10$ fluid simulations run by our group using the {\tt iharm} code~\citep{gammie2003}. The simulations have a characteristic correlation time of order $100$ light-crossing times for the event horizon (denoted $100 T$).  The simulations are run for $10^4T$ or $\simeq 100$ correlation times, yielding a set of $\simeq 1000$ independent physical realizations of the source.  

The synthetic images are generated using the {\tt ipole} code~\citep{moscibrodzka2018} with radiative transfer parameters drawn at random from the relevant parameter set, with a uniform distribution of snapshot times across the interval when the simulation is in a putative steady state.  In addition, the orientation of the image (determined by a {\em position angle} PA) is drawn randomly from $[0,2\pi)$, the center of the image drawn uniformly from $[-10,10]$ pixels in each coordinate, and the angular scale of the ring in pixel units is also drawn uniformly from a small interval corresponding to a $\pm10\%$ change in diameter.  The latter three variations were implemented to prevent the neural network from correlating source parameters with numerical effects like chance pixel alignments.   

Using our synthetic data set, we extend the analysis performed in~\citet{van2020deep} in several ways.  First, by training on both MAD (high magnetic flux) and SANE (low magnetic flux) models, we test whether our neural network can discriminate between simulations with similar spin parameters but different flux states; \citet{van2020deep} consider only SANE models. We also test our neural network on images produced from fluid simulations with spins that were not included in the training data set. Finally, we perform an analysis of feature maps to explore what information the neural network uses to make its predictions.

\section{Training}

\begin{figure}[ht]
\vskip 0.2in
\begin{center}
\centerline{\includegraphics[width=\columnwidth]{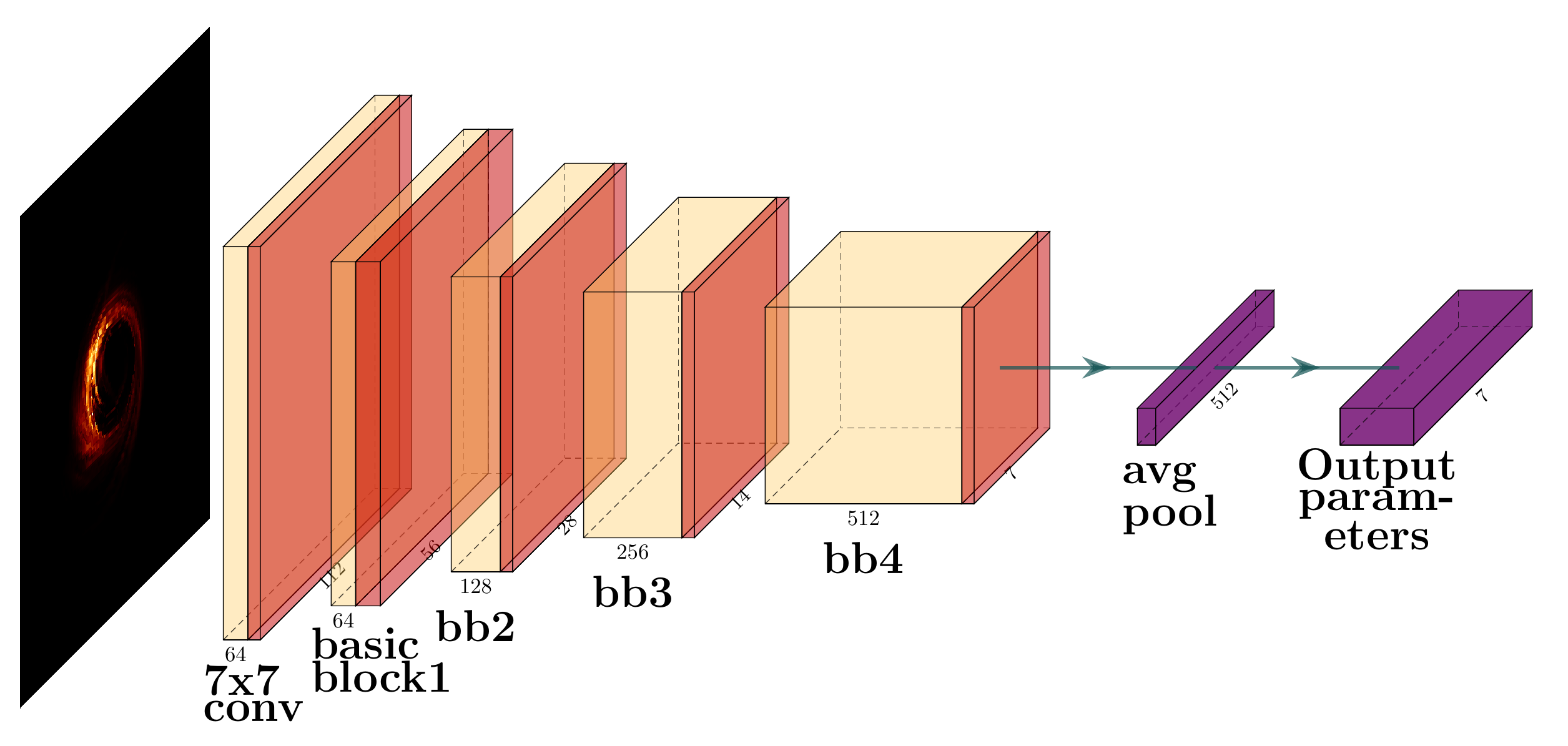}}
\caption{The modified \resnet architecture we use to classify black hole images. Each input image is fed into the network on the left side of the figure and passes through the vanilla convolutional and combined basic block (convolution + norm + rectifier) layers. Plot visualized by  \textit{PlotNeuralNet} package \cite{haris_iqbal_2018_2526396}}
\label{fig:Resnet-arch}
\end{center}
\vskip -0.2in
\end{figure}

We randomly select $200,000$ images from the million-image library and split that group equally into training and testing sets. For further validation, we construct three blind data sets. Data set A consists of $40,000$ images drawn from the million image library. Data set B is a set of $12,000$
images produced from the same GRMHD simulations that were used to construct the million-image library, but with different values for radiation physics parameters than were used to generate the training set. Data set C is composed of $1,100$ images produced from GRMHD simulations that were not used to train the network and that had different spins.

In this work, we build our neural networks with the Python 3 deep learning library \emph{pytorch} and use the standard deep residual network architecture \resnet for our base network topology~\citep{he2016deep}. 
Because the images we consider are monochromatic, we replicate the input data into each of the R, G, B channels.

We train two neural networks with this scheme. For parameter estimation, we use a network with a fully connected final layer that contains seven neurons spanning over the seven parameters that vary across the generative simulations. In this work, we only consider one of the seven parameters---the spin of the black hole---and ignore output from the other six neurons. We also train a classifier network to discriminate between MAD and SANE images using binary cross entropy loss.

Before training, we initialize the neural network weights from the pre-trained \resnet model.
For both classification and parameter estimation, we set the neural network training epoch to 200 and the batch size to 50. We train our network with a gradient descent scheme based on the Adam optimizer \citep{kingma2014adam} and set the default learning rate to $10^{-4}$.

\begin{figure}[ht]
\vskip 0.2in
\begin{center}
\centerline{\includegraphics[width=\columnwidth]{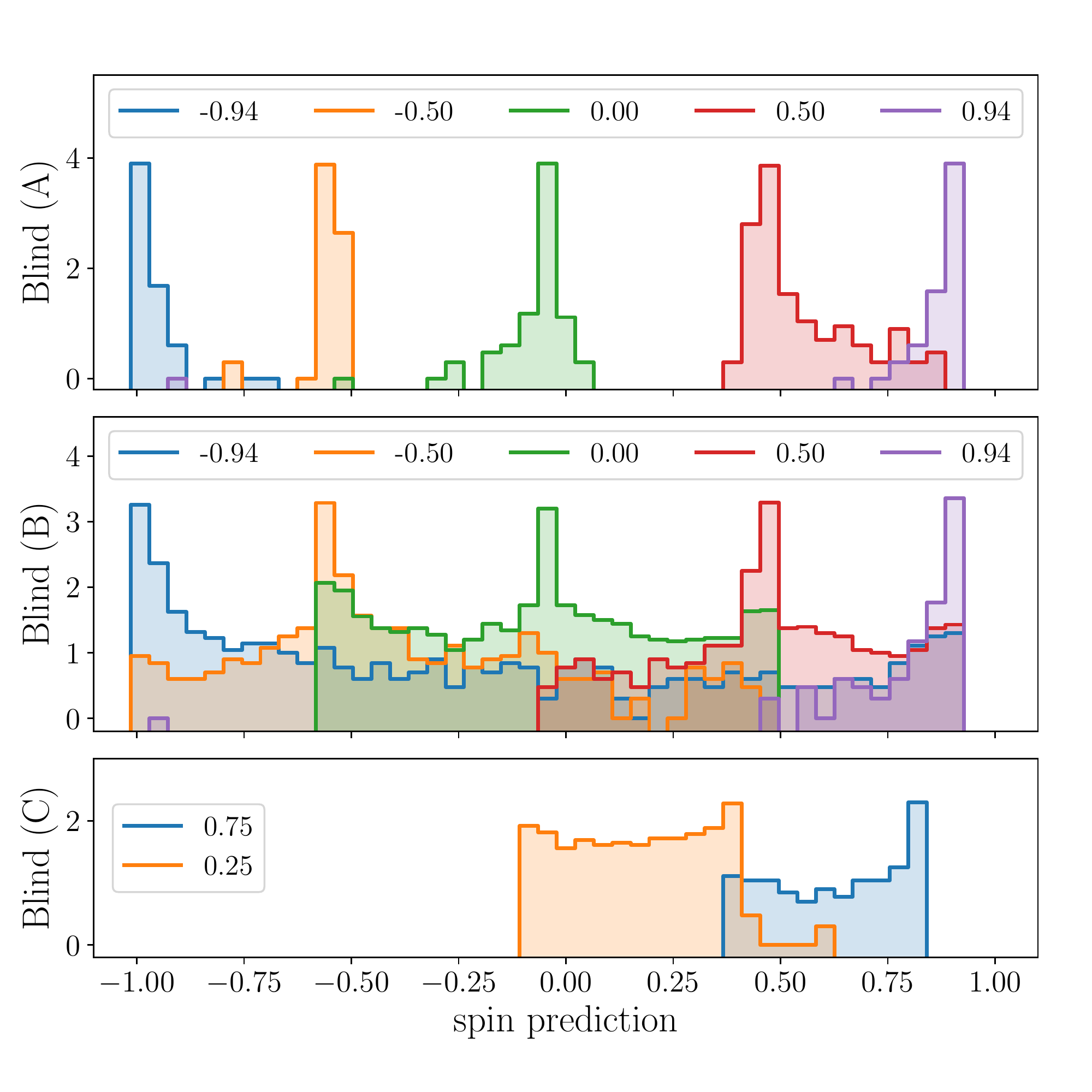}}
\caption{
Comparison of neural network spin prediction versus ground truth value for the blind data sets. Truth values are set by the fluid simulations and are one of $0, 0.25, \pm 0.5, 0.75$ or $\pm 0.9375$. Histogram bin heights correspond to the logarithm of the number of predictions within a given bin. 
}
\label{fig:ground-truth-comparison}
\end{center}
\vskip -0.2in
\end{figure}

The synthetic images are generated at a $160 \times 160$ pixel resolution where each pixel corresponds to 1 $\mu$as$^2$ on the sky. To interface the images with the pretrained \resnet architecture, we increase the resolution to $224 \times 224$ pixels using bilinear interpolation.

The MAD/SANE classifier network is able to correctly identify images from the three blind data sets A, B, and C, with $99.89\%$, $93.79\%$, and $98.33\%$ accuracy respectively. When predicting spin (ignoring other quantities), the regression network achieves standard deviations of $0.013$, $0.209$, and $0.164$ respectively versus the truth values. Data set C was constructed from two separate fluid simulations with different black hole spins. The mean prediction value for the truth $=0.75$ spin is $0.79$, and the mean prediction value for the truth $=0.25$ images is $0.24$.

Figure~\ref{fig:ground-truth-comparison} compares the estimated spin values for images in each of the three blind data sets. Because the images were generated from a small set of fluid simulations, the ground truth spin values are restricted to one of $0, 0.25, \pm 0.5, 0.75$ or $\pm 0.9375$. This discreteness was also present during the network training, and thus the network tends to favor these values in its predictions. This training--enforced prior drives the multi-modality observed in, e.g., the zero spin predictions for data set B, in which incorrect predictions were more likely to select the $\pm0.5$ values, versus intermediate ones.

For data set C, which was produced using images of fluid simulations that were not a part of the training data, the network favors spin values it has already seen; it tends to guess values for spin that are roughly consistent with the truth values. This is interesting because it suggests that the network may be keying on image features that map smoothly between different values of spin. Understanding such a mapping, if it exists, could help target future observation campaigns and technologies. 

\begin{figure}[ht]
\vskip 0.2in
\begin{center}
\centerline{\includegraphics[width=\columnwidth]{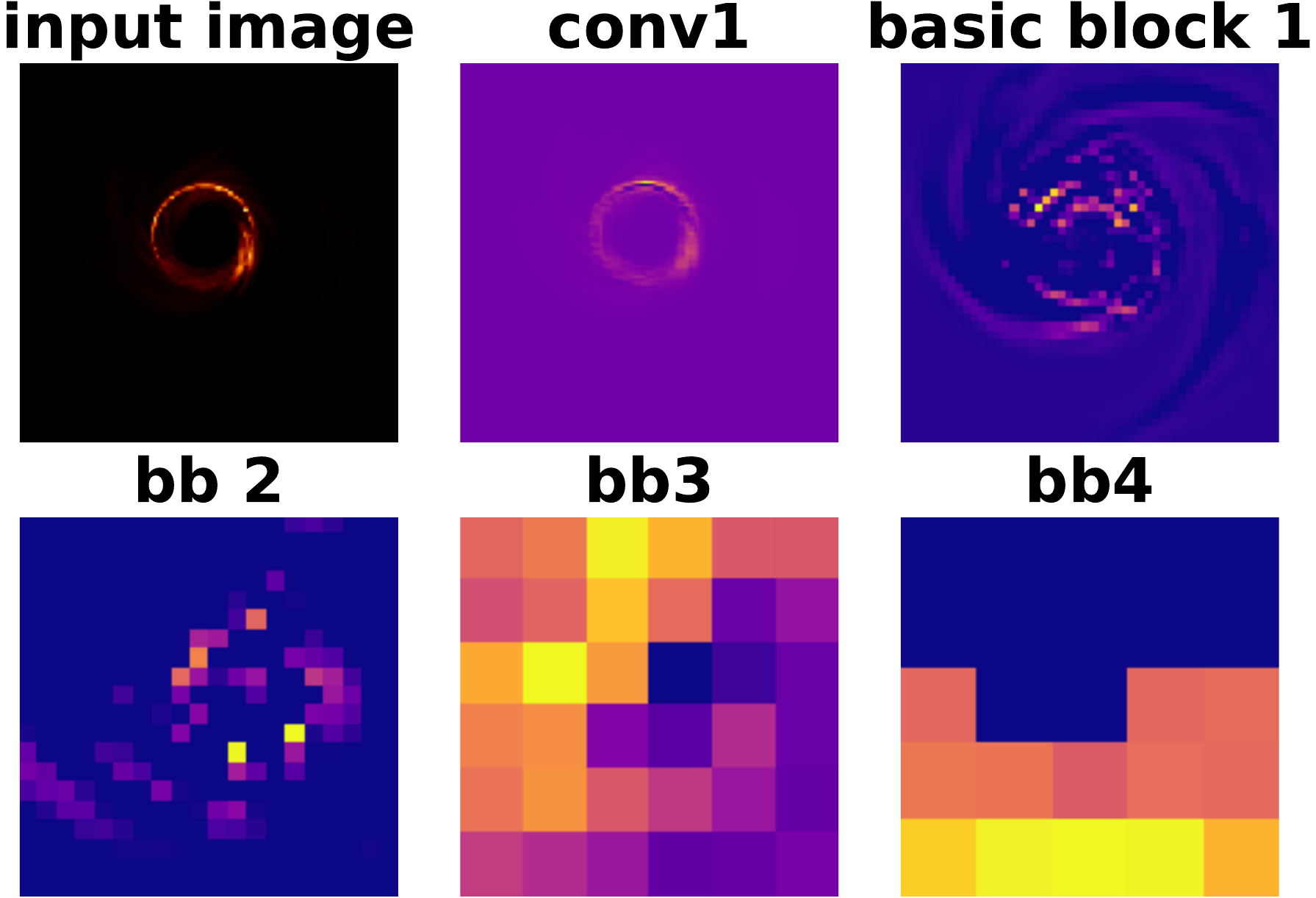}}
\caption{Representative feature maps for each of the five main \resnet layers for a synthetic black hole image. Later, lower resolution maps are often harder to interpret, whereas earlier maps are often quite similar to the input image. In this work, we explore the feature maps of the first basic block (bb) because they can be readily interpreted and they are not so similar to the input image as to be uninteresting. 
}
\label{fig:Feature map sample}
\end{center}
\vskip -0.2in
\end{figure}

\begin{figure*}[ht]
\vskip 0.2in
\begin{center}
\includegraphics[width=.45\textwidth]{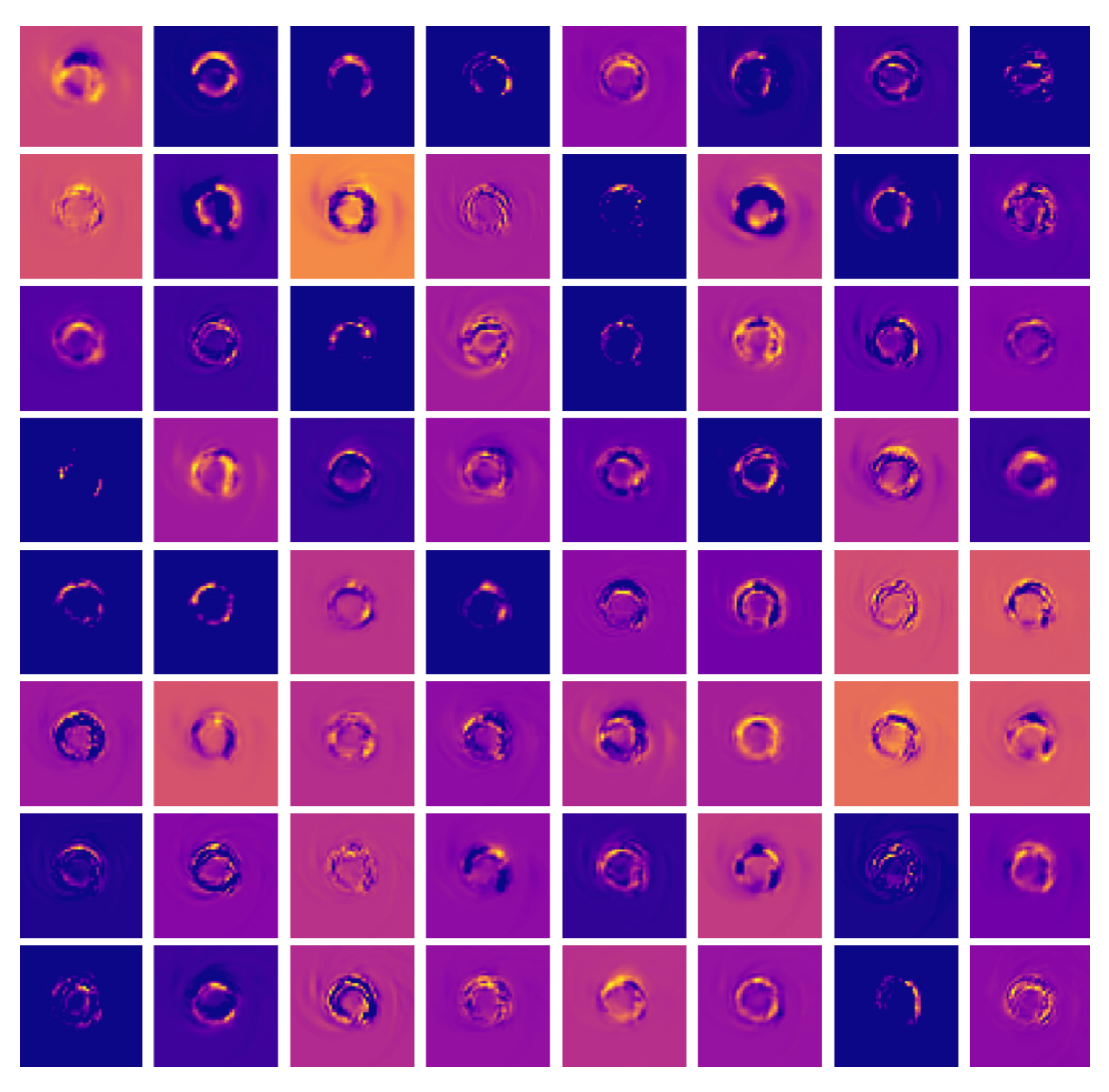} $\qquad$
\includegraphics[width=.45\textwidth]{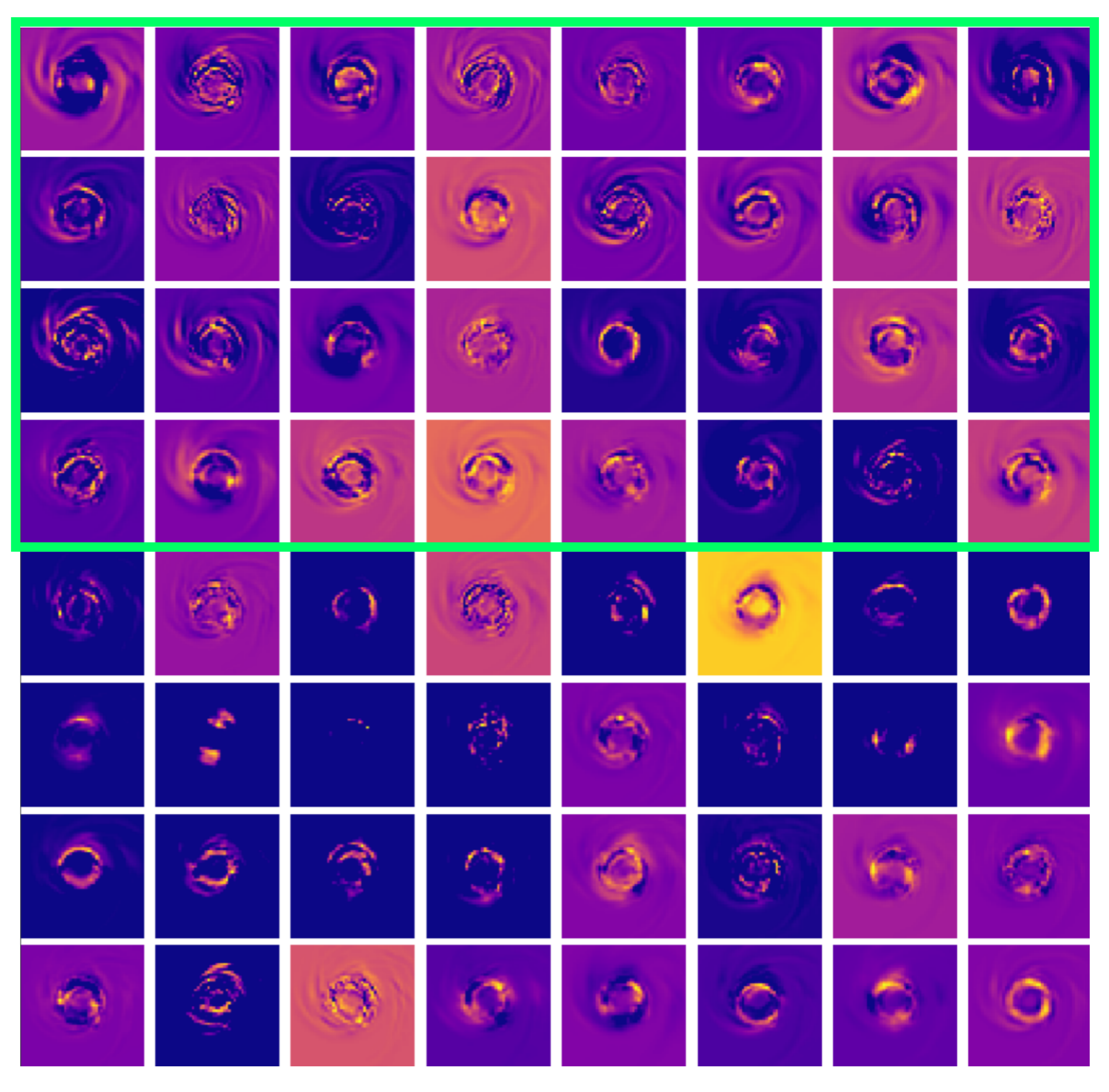}
\caption{
Feature maps for each output of the first basic block when fed trial image shown in Figure~\ref{fig:Feature map sample}. Left panel: feature maps produced from untrained vanilla \resnet network. Right panel: feature maps produced from trained network. Although both untrained and trained networks produce feature maps that identify the prominent ring-like structure in the image, the trained network seems to assign more relative importance to spiral features in the image. We sorted the feature maps so the ones with spiral features are on the top.
}
\label{fig:bb1-feature-map-comparison}
\end{center}
\vskip -0.2in
\end{figure*}

\section{Feature maps}

We use feature (or activation) maps to visualize how the neural network makes its predictions for spin. When fed a trial image, feature maps show the importance of different pixels within a convolutional layer with respect to the classification task. For visualization purposes, we subtract the mean map value from each pixel and then normalize pixel values by the standard deviation over the full map.

In Figure~\ref{fig:Feature map sample}, we present representative feature maps from each of the \resnet layers shown in Figure~\ref{fig:Resnet-arch}. Each layer in the network provides different information. In this work, we focus on feature maps from the first basic block because we find that they are easier to interpret compared to maps from subsequent layers, yet are still different enough from the input image to be interesting.

To understand which features the network learned during the training procedure, i.e., which features are both characteristic of black hole images and useful in differentiating between black holes with different spins, we compare the feature maps produced by the same trial image run through either an untrained vanilla \resnet network or the same network after it has been trained. The full feature maps of the first basic block for each network are shown in Figure~\ref{fig:bb1-feature-map-comparison}. 

One significant visual difference between the untrained and trained feature maps is that the maps produced by the trained network seem to pick out low brightness spiral arm features in the images. The physical origin of these spiral arms is complicated and related to turbulent features in the black hole accretion disk and jet. It is unsurprising that jet-bound image features may heavily inform the spin classification, since the jet is often thought to be connected to the spin of the hole (\citealt{Blandford1977}; \citetalias{EHT5}).

\section{Discussion}\label{sec:summary}

We have shown that it is possible to use a version of \resnet 
to accurately infer black hole spin and magnetic flux values from a library of synthetic images. An analysis of feature maps suggests that the neural network is keying---at least in part---on low surface brightness features. It is possible that these features arise from stochastic, slowly evolving structures in the fluid simulations that remain present over the full sample of images, and that the neural network is therefore overfitting the data and learning the base fluid simulations rather than features of the image morphology; however, we note that the network was able to classify images of fluid simulations that  were not part of the training set.  We plan to explore the overfitting problem in future work. 

If the network is primarily using the low-intensity spiral arm features to perform the classification/regression, it is possible that classification/regression will be significantly harder when attempted with real world data. This is due to both the low signal-to-noise ratio in the data, which often obscures the spiral arm features, and the fact that the true data consist of Fourier components, not images. Since real world Fourier domain coverage is sparse, it is possible that small features cannot be recovered (or are significantly perturbed) in the data.

Our synthetic image set was produced with an angular resolution higher than what is currently possible with the EHT, and our synthetic image data do not account for observational error, atmospheric phase fluctuations, or thermal noise. We plan to address these shortcomings in future work.
We also plan to work directly with synthetic sparse Fourier-domain data. Since the EHT captures information about source polarization and time dependence, we plan to study whether these additional data can further constrain physical parameters of the source. We will apply feature extraction techniques  \citep[see, e.g.,][]{ghorbani2019towards} to gain a more quantitative understanding of the features. It is critical to understand the origin of accurate neural network classifications/regressions so that we can be confident that we are not overfitting. 

\section*{Acknowledgements}

The authors would like to thank Katie Bouman, Been Kim, Asad Khan, and Derek Hoiem for useful discussions. The authors also thank the reviewers for useful comments.

JL, GNW, BP, and CFG were supported by the National Science Foundation under grants AST 17-16327 and OISE 17-43747.  GNW was supported by a research fellowship from the University of Illinois. 

This work used the Extreme Science and Engineering Discovery Environment (XSEDE) resource stampede2 at TACC through allocation TG-AST170024.  This work utilizes resources supported by the National Science Foundation’s Major Research Instrumentation program, grant 1725729, as well as the University of Illinois at Urbana-Champaign. JL thanks the AWS Cloud Credits for Research program. JL and CFG thank the GCP Research Credits program.

\bibliography{bib}
\bibliographystyle{icml2020}

\end{document}